\documentclass[allclo,superscriptaddress,eqsecnum,amsfonts,showpacs]{revtex4}
\usepackage{epsfig,rotating}
\usepackage{feynmf}

\newcommand{\be}{\begin{equation}}
\newcommand{\ee}{\end{equation}}
\newcommand{\bea}{\begin{eqnarray}}
\newcommand{\eea}{\end{eqnarray}}

\newcommand{\ep}{i\varepsilon}
\newcommand{\nn}{\nonumber}

\begin{document}

\preprint{ \parbox{1.5in}{\leftline{hep-th/??????}}}

\title{Vertical mass hierarchy in the case study of a hybrid model}

\author{Vladimir ~\v{S}auli}
\affiliation{DTP INP Rez, CAS. }


\begin{abstract}
We  study a hybrid model in which the Technicolor and fundamental heavy scalars live together.
We concern extreme case, where electroweak symmetry breaking (EWSB) comes almost entirely from the  strong dynamics of  Technicolor, 
but the masses of  the light fermions, e.g. leptons arises due to the weakly interacting scalars with one universal Yukawa interaction.
We have found that  $SU(2)$ scalar two doublets, which are private but universal for each generation,   
offer a natural explanation of the vertical mass hierarchy. There  is no large Yukawa coupling in the theory and the scalar masses are expected to be quite universal and large. The lepton mass hierarchy arises  radiatively  from the  quark mass  hierarchy, 
which is affected by the scalar sector as well. Lepton and down quark masses are calculated in a simplified case of neglected mixing.

\end{abstract}

\maketitle

\section{Introduction}
\begin{figure}[b]
\centerline{\epsfig{figure=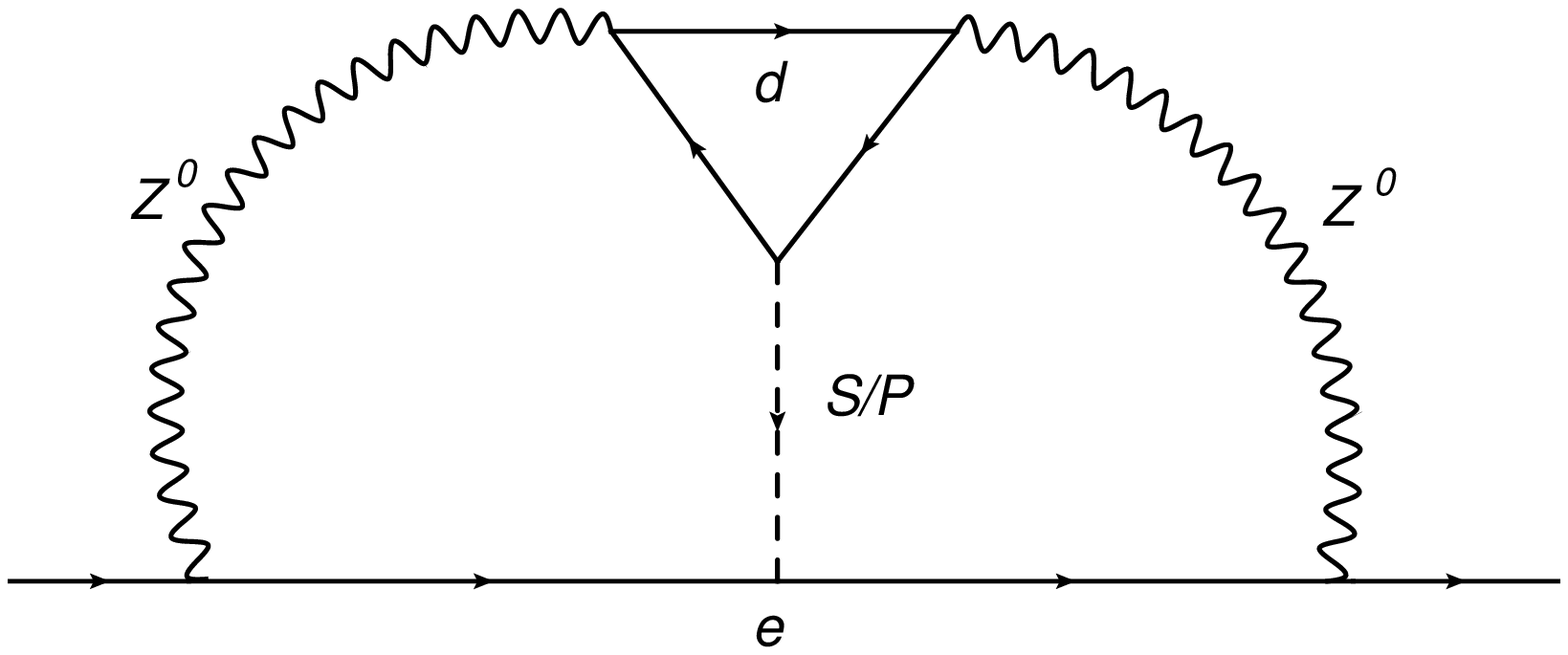,width=6truecm,height=3truecm,angle=0}
\hspace*{1.0truecm}
{\vspace*{3.0truemm}\epsfig{figure=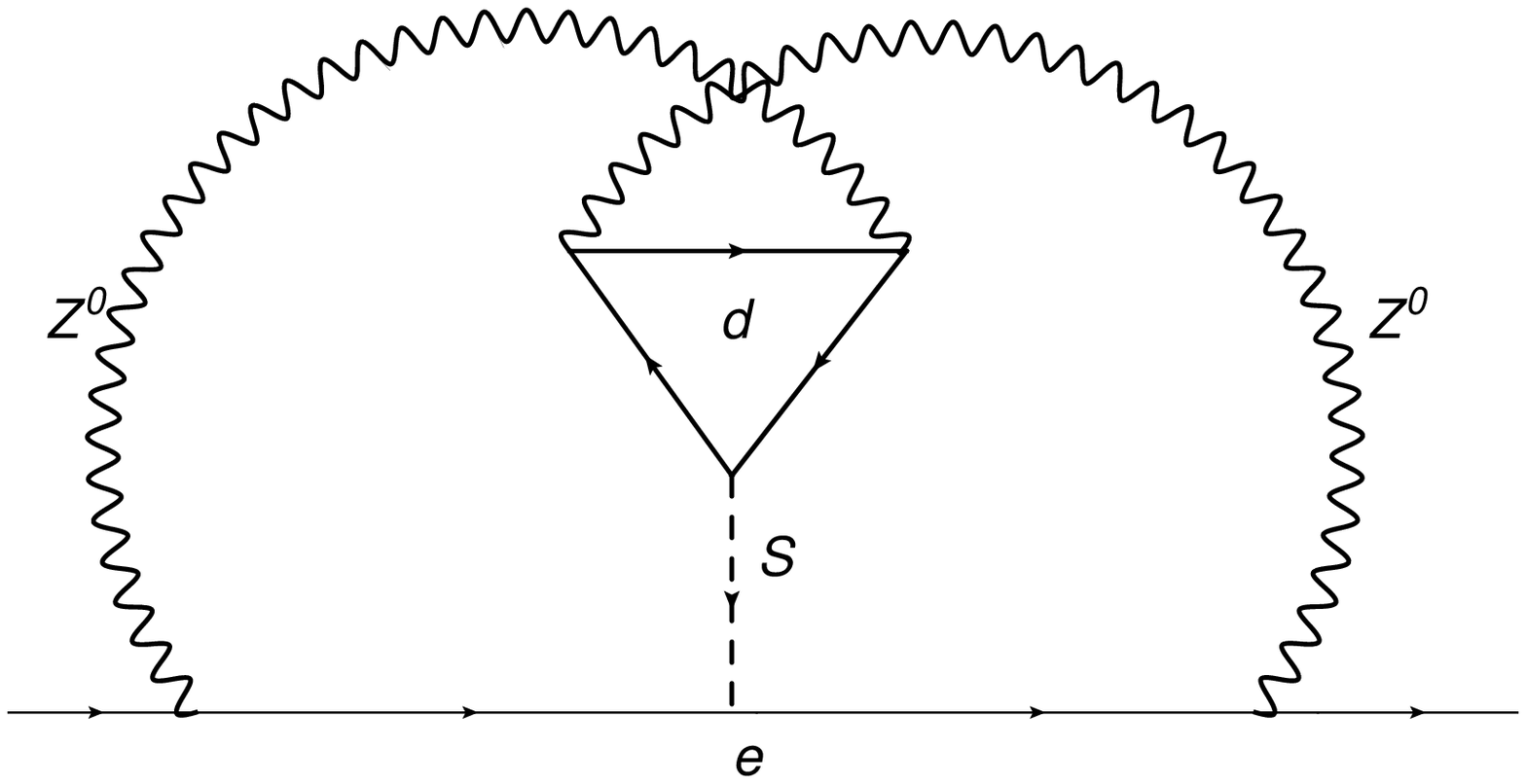,width=6truecm,height=3truecm,angle=0}}}\vspace*{-0.4truecm}
\caption[caption]{The examples of fountain diagrams which contributes non-trivially to charge lepton masses,  momentum labeling is shown explicitly.}
\label{explicit}
\end{figure}

Up to an incoming new 126GeV LHC scalar, confirmed and well established part of the Standard model is  presence of the  three fermion generations interacting with  gauge fields. Four gauge vectors interact rather weakly with matter of families with  couplings strengths $g_{SU(2)}\simeq g_{U_Y}\simeq 0.1$  . At very low energies, the remaining eight gauge fields - the gluons- interact strongly, however  with only quark  family  members.
According to  observed  heaviness, the SM leptons and fermions has been divided into the three families. In  the SM, the problem of the mass hierarchy is solved by phenomenological successful setup of  the Higgs Yukawa couplings. Contrary to the pure gauge couplings,  Higgs Yukawas are necessarily non-universal ranging their numerical values from $y_\nu\simeq 10^{-14}$ for neutrinos to  $y_t \simeq 1$ for top quark and for purpose introduced negative mass term of the Higgs field  turns to be responsible for EWSB and for the generation of the  lepton and the current quark mass generation in the SM.

For a light Higgs, the radiative "corrections" from heavy quark loops  are proportional to the relatively  large Yukawa couplings $y_q$. When introducing suitable ultraviolet regulator $\Lambda \simeq \Lambda_{GUT}$, these corrections appear to be very early larger then the electroweak scale breaking $v_{SM}=247 GeV$. For a single Higgs field the correction to the Higgs field mass $\delta v \simeq \delta m_{Higgs} \simeq |y_q| \Lambda >> v $ as a well  known consequence of unprotected SM against the presence of  quadratic divergences associated with the presence of scalar fundamental field in the theory. Apart of this so called large hierarchy problem, an unexplained hierarchy of Yukawa couplings remains mystery as well.
 
  The  supersymmetry is the most popular idea, which when realized in the Nature can solve the problem of naturalness. As there is no 
evidence for the existence of the supersymmetric partners at the experiments, it must be "softly" broken and the susySM appears to be unnaturally  constrained at least in its minimal version.  Another  possibility is the  absence of a light fundamental scalars in the spectrum. We assume that such scenario can really happen in the nature, however we do not exclude fundamental scalars completely. Instead off, we assume the main part of EWSB
stems from strongly interacting sectors of Technicolors \cite{FASU1981,WE1976,SU1979,EILA1980}, while the scalars are  heavy   that the large hierarchy problem and naturalness is  largely ameliorated. To achieve this, all the Yukawas should be small, the scalars should be quite heavy $M>10 TeV$ and their number is limited in a way they cannot be  predominantly responsible for EWSB and their effect mostly  decouple the scalars from low energy observables.
In pure Technicolor scenario of EWSB the Higgs is composite and the lightest scalar corresponds to a bound state of Techniquark-antiquark pair, i.e. Technimeson. In hybrid models like here, the light composite Higgs can mix with the light fundamental unconstrained scalars as suggested in the recent paper \cite{SONI2013} with the price that the large quadratic divergences associated with the presence of light fundamental scalar are back into the game.  In the case of SUSY Technicolor \cite{SAM1990,DOB1995,ACST2011,COKUPEPE2012} presence of scalars multiplets reappear to be natural. At this place, one could mentioned that when introducing scalars into the theory, Supersymmetry is  certainly not the only possible  way which can make the theory ultraviolet complete. The higher field derivatives at some higher scale  makes the model  free of quadratic divergences as well, see for instance 
\cite{LEWI1969,GRCOWI2008,CALE2008,WUZH2008,WUZHII,ROSH2009,INKW2010}. In our bottom-up study we do not consider ultraviolet completion, instead of we construct a part of such model, which is predominantly responsible for a mass hierarchy only and where the problem of naturalness is transfered to much higher then in the case of the SM. At last but not at least , let as mention that non-susy the strong Technicolor-like dynamics of very heavy scalars remains natural as well \cite{benes,smetana}.

The presence of heavy scalar can  effectively supply, if not fully replace, the  Extended Technicolor mechanism. Actually, the radiative corrections mediated by heavy scalars appear as a  good  candidate for an explanation of  the observed mass pattern.
 Note for completeness, the attempts to  calculate  quark and lepton masses from the radiative corrections  have long history \cite{GEGL1973,IB1981,BAKAMO1988,BA1988,BAMA1989,BA2007,BOFO2008}. 

In Technicolor scenario the strong dynamics breaks the chiral symmetry through the mechanism quite similar to QCD. Three massless technipions modes are then eaten by longitudinal polarization of electroweak vector bosons $W$'s and $Z$.
A correct values of boson masses $M_{z,w}$  require an increase of the sliding scale of the theory to be $\Lambda_{TC}\simeq  TeV$ (here $\Lambda_{TC}$ is the analogue of $\Lambda_{QCD}$). Apart of EWSB one expects plethora of technihadrons on the physical spectrum.

 To  avoid strong conflict with the experiment, the masses of SM fermions must be generated more indirectly. One possible way is the idea of   Extended Technicolor  where the SM fermions are embedded into the representation of ETC group. It is worthwhile  to mention here, such theory is  ultraviolet complete (it is free of quadratic divergences) and can lead to the observed quark mass ratios and mixing angles \cite{APPIRS2003}. Recently observed LHC 126 GeV boson by CMS \cite{CMS} and ATLAS \cite{ATLAS}  collaborations was suggested  \cite{MAYA2012,GKRS2012,CCILMRS2012,COKUPEPE2012}
 to be  a pseudogoldstone boson of broken gauge symmetry,.i.e. one of the Technipions of various  Technicolor based models. However note, the composite CP-even scalar is surviving possibility as well \cite{FTS2012,FOFR2012,NATALE2013} (in this case all technipions should be  absorbed).

In this paper we propose an alternative  solution which combine TeV-scale Technicolor  EWSB with a presence of fundamental heavy scalars partially responsible for mass hierarchy of the SM fermions.   We assume that the strong coupling  Technicolors is predominant in the heavy quark sector, while the values of the light quark masses can be  affected by quantum corrections with  heavy scalars scalars as well.   Furthermore the leptons are not embedded in any ETC group representation, the get their masses purely from the loops with scalars, however combined with the loop of already massive quarks. As a result of an example model considered in this paper we will saturate most SM masses by using single heavy mass scale and single universal Yukawa. The only heavy upper quarks $c$ and $t$ have origin in  the strong dynamics and  their closed loops will  serve as a seed for the all other fermion masses in the original SM.

 For this purpose we introduce two doublet  -opposite Weak hypercharged- scalars. As we assume they do not participate to much on EWSB we do not assume wrong sine of the mass term , avoiding thus condensation at classical level. 
Furthermore, to achieve first hint for the mass hierarchy and in order the avoid phenomenologically unacceptable large neutral current we introduce these two pair of heavy scalar for each generation  separately.  

The leptons then become become massive through the  3loop "fountain diagram" depicted in  Fig.1 and Fig2, where the quark closed loop  is presented in. Thus, in this  model one can roughly say that the lepton mass hierarchy comes almost entirely from the hierarchy of quarks. Evaluating a multiloop diagrams we present simple rules for the ratio of lepton masses as a function of quark masses at Technicolor and heavy boson  scale.

\begin{figure}[right]
{\epsfig{figure=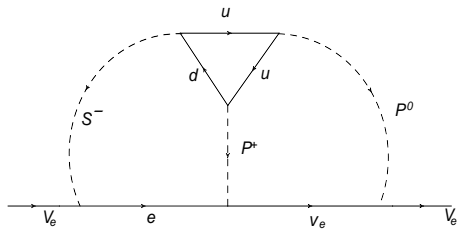,width=6truecm,height=3truecm,angle=0}} \vspace*{-0.4truecm}
\caption{An example of fountain diagram responsible for neutrino mass generation.}
\end{figure}

 The paper is organized as follows: In the next Section we write down the Lagrangean of the Model. In the Section III. we mention some important properties of Green functions in  Technicolor model. In the Section IV. we calculate the dominant contributions to neutrino masses. A more rough estimate is presented for SM charged leptons in the Chapter V. We discuss and conclude in the Section VI.  

\section{Decoupling  of heavy scalars from EWSB and electroweak oblique corrections}

In the SM this is the scalar doublet -The Higgs- which enforced to condense at classical level, which  gives rise the masses of all other particles.
In presented model the role of heavy scalars is not to break EW symmetry down to the electromagnetic group, 
but instead off, the scalar sector replaces the Extended Technicolor mechanism in the leptonic sector.
Since the study of such possibility is the main aim of presented paper we have to identify conditions which allow natural
implementation of the heavy scalar doublets back into the game.

First of all, any new scalar field can condense, if not at tree level, the through the raditaive coorections. If  condensates are formed they  can potentially contribute to the EWSB. These condensates,  when summed up with the contribution of TechniHiggs must  end up with complete value $v=246 GeV$.  

The second problem can arise if the introduced electroweak doublets make an observable changes. Note, an adding of a number of EW active scalars
back into the game can lead to potential changes in formulas for various experimental observables. Furthermore the same is valid for Technicolors, where the especially the number of EW active new fermion doublets is highly limited. To get rid of this problem we will consider limiting case in which case 
the scalar sector decouple and will not contribute to the electroweak precision observables at all. 
In this section we will show that if the extension of SM consist from enough  heavy electroweak scalars which do not condense at classical level, they effectively decouple from the both- form the  EWSB and as well as from electroweak precision observables.

Yukawa interaction with a heavy neutral scalar will be an important part of the model. 
As the heavy quark masses are generated at (Extended) Technicolor scale $\Lambda_E$  the  tadpole diagrams with closed quark  loops contribute to the effective potential at one loop level. To find true ground state of the theory one have to  minimize the effective potential, which requires solution of  cubic "tadpole" equation in general. 
However , for a small coupling one can  use the perturbation theory, which leads to the  condensate value   $<\phi>=v=-\lambda/M^2$ , where $\lambda$ is a sum of tadpoles diagrams and $M$ is a heavy mass of the scalar. Recall the tadpole would be  zero for a massless quarks (for Technicolor and QCD switched off).

The  one loop tadpole diagrams   read
\be
\lambda=-4N_c y\int^{\Lambda_E}\frac{d^4k}{(2\pi)^4}\frac{m_t}{k^2+m_t^2}
+24\lambda_4 v \int^{\Lambda_E}\frac{d^4k}{(2\pi)^4}\frac{1}{k^2+M^2} \,
\ee
where we have shown  the contribution from the heaviest  quark  mass $m_t$, $y$ is Yukawa coupling constant and 
the second term is in fact second order contribution as it  appears due to the shift already generated,  note $v \lambda_4$ is triliner scalar if  $ \lambda_4$  is quartic one.

In words, this  the heaviness of scalar which prevent large condensation of the scalar field and effectively  decouple heavy 
scalars from low energy physics.  Assuming for a while, that the heavy scalar mass replaces the scale of  non-abelian Extended Technicolor we assume UV divergences are regulated at $\Lambda_E\simeq M$ . Thus for small  Yukawa $y$ we can estimate
\be
v=\frac{3y}{2\pi^2} m_t(M)
\ee
While the top quark mass can  be still  large at the scale $M$, it is the smallness of Yukawa coupling, which  avoid a large contribution to EWSB. We can
anticipate here, that all scalars  presented in the model considered in the next section  
participate on EWSB by negligible amount  ($\simeq 1$ of $246$ $GeV$).  

More interesting from mass hierarchy perspective,  there is a cancellation mechanism for higher point vertices. This  prevents further enhancement of symmetry breaking by  a tadpoles. Note for completeness, that in the models where symmetry breaking appear at classical level, the tadpoles can be effectively ignored as they can be effectively summed and absorbed through the renormalization \cite{Collins,KUMU1996}. In a theory where  symmetry is broken dynamically, the classical mass is not presented.  Thus to deal with renormalizable theory, the corrections which generates physical masses must be  finite, and all allowed infinites should be absorbable  by the model constant presented in the unbroken theory- mass of the scalar and Yukawa coupling in our case.

\begin{figure}[right] \label{tadpole}
{\epsfig{figure=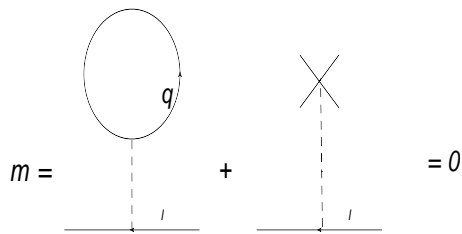,width=6truecm,height=3truecm,angle=0}} \vspace*{-0.4truecm}
\caption{Diagrammatical representation of one loop tadpole cancellation in Yukawa model.}
\end{figure}

For our purpose we will  exhibit the cancellation of fermion mass at one loop level. We have shown a nonzero vacuum expectation value due to the one loop tadpole. This,  when inserted to the Yukawa interaction, generates a current or constant mass of any fermion, which inteacts with the scalar, i. enot only quarks.  Actually such  contribution is exactly cancelled by the tadpole attached to the external fermion line. Note the importance of the extra minus sign due to the quark closed loop. Graphically, the one loop cancelation is shown at Fig. (\ref{tadpole}).

As the chiral symmetry in the lepton sector is broken in a quite complicate manner we are not awared about a nonperturbative and/or  the all orders proof,  therefore we do not assume the aforementioned mechanism persist to the all orders of theory. However, regarding the mass generation of the  leptons,  the main role of the tadpoles is to cancel each other, which is certainly true in lower order perturbation theory and we will ignore tadpole by considering only 1-particle irreducible selfenergy diagram only.


A number of electroweak observables 
can be potentially affected by the radiative corrections generated by a novel loops with non-SM field content.
These changes are nowadays conveniently summed up into  the few  electroweak oblique parameters \cite{PETA1990}.
Here we simply quote that heavy scalar with equal heavy weak izospin component decouple from the electroweak oblique corrections. 
For the scalar doublet with general isospin mass splitting the result reads \cite{LI1995,ZHYALI2006}
\bea
 S&=&-\frac{Y}{3\pi}\log{\frac{M_{u}}{M_{d}}}
\nn \\
 T&=&\frac{1}{12\pi s_w^2 c_w^2 M^2_z}\frac{(M_u^2-M_d^2)^2}{(M_u^2+M_d^2)^2} \, ,
\eea
where we have used conventions established in  \cite{ZHYALI2006},  $Y$ is a weak hypercharge, $s_w$ and $c_w$ is shorthand notation for sin and cos of Weinberg mixing angle respectively, indices $u,d$ stand for upper and down component of $SU(2)$ doublet.

Obviously, the presence of heavy degenerate scalar is not excluded by the current electroweak precision fit, thus in our hybrid model the value of $S,T..$ parameters are fully determined by the details of strong coupling Technicolor theory only \cite{survivor}.

\section{Private lepton-quark four doublets model}
          
The main aim of our model is a natural explanation of lepton mass hierarchy within a minimal introduction of a new  complicated field content and couplings.
For this purpose we introduce  two new private (individual for each generation) scalar EW doublets $N,S$ with SM group assignment $(2,+1/2)$ and $(2,-1/2)$. In this way, the SM  is  extended by a six new doublets. While general Yukawa and various scalar couplings can lead to chiral (and parity violating) interaction we can finish with physical spectrum with neutral and charged scalar and pseudoscalar Yukawa interactions only. Thus just one type of scalar interacts with given generation quark and lepton content and  the flavour changing neutral current is avoided by the privacy of scalars.
Thus these are Yukawas which tell us which lepton doublet matches a given quark doublet. 
 Furthermore, we assume all Yukawas are universal for the quarks and the leptons  and the  numerical value is identical for the all three generations. t
The assumed  coupling strength is relatively small $\alpha_{y}=\frac{y^2}{4\pi}\simeq 10^{-3}-10^{-4}$ and the scalar self-interaction of introduced field content is constrained  further by a discrete symmetry. In presented  study of the model we neglect the intergeneration mixing 
(in all subsectors), leaving thus CKM matrix simply diagonal. Note, the  mixing can be accomplished in  many various  ways, while 
the main features of presented model are kept intact, e.g. the

1. The Yukawas are universal and small.

2. no large  CP violation is introduced through the scalar sector. 

3. The smallness of neutrino masses should not affected by introduced mixing. 

We also add right neutrino singlets. Assumed  ETC interaction, which we do not specify in details breaks to the Technicolor subgroup  and SM, thus required symmetry is $SU_{TC}(2)*SU_w(2)*U_Y(1)$  at the scale $\simeq$ 10 TeV. 
Further we assume that condensation of Techniquarks (with a possible participation of top quark)  is almost entirely responsible for 
electroweak symmetry breaking. Leptons are not embedded into any representation of  ETC group, therefore they  communicate with $TC$ sector only  indirectly. Consequently they receive their masses through the quark loop due to the presence of scalar doublets. For both -charged leptons and neutrinos as well- first nontrivial contribution will appear at three loop level.


Our model, which is UV complete up to the scale  $\simeq 1000 TeV$,  can be described by the following effective Lagrangean
\bea  \label{model}
L&=&L_{SM-Higgs}+L_{TC}+L_{S+Y}
\nn \\
{\cal L}_{S+Y}&=&L_e+L_{\mu}+L_{\tau}+...
\eea
where $L_{SM-Higgs}$ is the Standard Model Lagrangean without Higgs; $L_{TC}$ denotes TC Lagrangean  and the rest represents Lagrangean of $2 N=n_f$ ($N=3$ for SM) scalar doublets weakly coupled to all SM fermions through the universal Yukawa interaction. We have labeled  intragenerational interaction by its charge lepton content, i.e. $L_{\mu}$  contains its own private scalar doublets $N_{\mu},S_{\mu}$ and three dots stands  for  possible forth   generation.
Using two complex doublets 
\be
N=\left(\begin{tabular}{l}
 $N^o$\\$N^-$
\end{tabular}\right)\, ;S=\left(\begin{tabular}{l}$S^+ $ \\ $S^o$ \end{tabular}\right)
\ee
then the gauge symmetry  allows to write down  general form of the electron private part of the  Lagrangean in the form  
\bea \label{lag}
\nn \\
{\cal L}_e&=&  D^{\mu}N_e^{\dag} D_{\mu}{N_e} + D^{\mu}S^{\dag}_e D_{\mu}{S_e}
M^2 N_e^{\dag}N_e-M^2 S_e^{\dag}S_e-\frac{\lambda_N}{2}\left(N_e^{\dag}N_e\right)^2
\nn \\
&-&\frac{\lambda_N}{2}\left(N_e^{\dag}N_e\right)^2
-\frac{\lambda_S}{2}\left(S_e^{\dag}S_e\right)^2-\lambda_{NS_1}\left(N_e^{\dag}N_e\right)\left(S_e^{\dag}S_e\right)
-\lambda_{NS_2}\left(N_e^{\dag}S_e\right)\left(S_e^{\dag}N_e\right)
\nn \\
&-&y_{\nu}(\bar{E}_L \nu_{eR}+\bar{U}_L u_R) N_e + y_e(\bar{E}_L e_R+\bar{U}_L d_R) S_e +h.c
\nn \\
&-&y_{\tilde{\nu}}(\bar{E}_L \nu_{eR}+\bar{U}_L u_R)\tilde{S_e}  + y_{\tilde{e}}(\bar{E}_L e_R+\bar{U}_L d_R) \tilde{N_e} +h.c
\nn \\
\eea
with SM antifermion left doublets $\bar{E}_L(\bar{e},\bar{\nu}_{e})_L, \bar{U}_L=(\bar{u},\bar{d})_L$ and  fermion right singlets $e_R,\nu_{e R},u_R,d_R$, as usually 
$\tilde{S}\tilde{N}=i\sigma_2 N^{c}$.

  Bare Lagrangean fermion mass is prohibited by gauge invariance in the  model, however as usually, the   physical  mass is just selfenergy evaluated at the same  scale. This fermion selfenergy is generated through the chirality violating Feynman graphs, thus in order to achieve parity conservation in QED dominated processes the  Yukawa and scalar couplings should be further constrained.  Recall here, that the general form of the radiative corrections to the fermion selfenergy   reads in momentum space $\Sigma=\not p (P_L a_1(p^2)+P_R a_2(p^2))+1(P_L b_1(p^2)+P_R b_2(p^2))$, where $P_{L,R}=1/2(1\mp\gamma_5)$. The lepton  fermion selfenergy is generated through the chirality violating Feynman graphs, thus in order to achieve parity conservation in QED dominated processes the  Yukawa and scalar couplings should be further constrained.
For this purpose, the universality of Yukawa coupling is enough condition, explicitly written:
\be
y_e=y_{\tilde{\nu}} \, \, ; \, \, y_{\tilde{e}}=y_{\nu}
\nn \\
y_e=-y_{\nu}\, .
\ee
Further we take $S^+_{new}=iS^+$ requiring that the non Yukawa part is invariant under this discrete symmetry (so $ \lambda_{NS_2}=0$). Doing so explicitly all Yukawas can be collected into  two terms of scalar charged an neutral current
\bea
{\cal L}_{(c.c.Y)}&=&-y\left(\bar{\nu} e N^{+} + \bar{e}\nu N^{-}+i\bar{e}\gamma_5 \nu S^-+\bar{\nu} \gamma_5 e S^{+}\right)
\nn \\
{\cal L}_{(n.c.Y)}&=&-y\left[\bar{e} e (s_R+n_R) + \bar{\nu}\nu(s_R-n_R) 
+i\bar{e}\gamma_5 e(s_I-n_I) -\bar{\nu} \gamma_5 \nu (s_I+n_I)\right) \, ,
\eea
where indices $R,I$ label the real and imaginary parts of the doublets neutral components. Similar can be written for quarks. 

Further we conventionally introduce  electrically neutral scalars and pseudoscalars through
the linear combination of $S$ and $N$ components as 
\bea
s_R=\frac{1}{2}({\cal S}_e+{\cal S}_{\nu})\, \, ; n_R=\frac{1}{2}({\cal S}_e-{\cal S}_{\nu})
\nn \\
s_I=\frac{1}{2}({\cal P}_e-{\cal P}_{\nu})\, \, ; n_I=\frac{1}{2}({\cal P}_e+{\cal P}_{\nu})
\eea

Thus the final form of our Lagrangean can be written in the following form
\bea \label{clasic}
\nn \\
{\cal L}_e&=& L_{kin}+\frac{M^2}{2}\left({\cal S}_e^2+{\cal S}_{\nu}^2+{\cal P}_e^2+{\cal P}_{\nu}^2+2S^+S^-+2N^+N^-\right)+V_{self}({\cal P}^2,{\cal S}^2)
\nn \\
&-&y\left(\bar{\nu} e N^{+} + \bar{e}\nu N^{-}+i\bar{e}\gamma_5 \nu S^-+\bar{\nu} \gamma_5 e S^{+}\right)
\nn \\
&-&y\left(\bar{e} e {\cal S}_e + \bar{\nu}\nu {\cal S}_{\nu}) 
+i\bar{e}\gamma_5 e{\cal P}_e  -i\bar{\nu} \gamma_5\nu {\cal P}_{\nu} \right) 
\nn \\
&-&y\left(\bar{u} d N^{+} + \bar{d} u N^{-}+i\bar{d}\gamma_5 u S^-+\bar{\nu} \gamma_5 e S^{+}\right)
\nn \\
&-&y\left(\bar{d} d {\cal S}_e + \bar{u}u {\cal S}_{\nu}) 
+i\bar{d}\gamma_5 d{\cal P}_e  -i\bar{u} \gamma_5 u {\cal P}_{\nu} \right) 
\, ,
\eea
where the scalar potential $V_{self}$, which does not include $\lambda_{NS_2}$, is not shown for brevity.

From the non-diagonal entries of SM Cabibo-Kobayashi-Maiani matrix and from the neutrino oscillation experiments we know that  
fermions  mix across the families.  In presented model the mixing of quarks is allowed, however  they enter always in the isolated closed loops which does can not introduce a mixing in  the lepton sector.  The origin of neutrino mixing must be therefore different. Recall  here,  that  the so called  private Higgs model  has been  already used in order to  explain mixing and  mass hierarchy in the SM \cite{MA2007,MAb2007,HSS2009,BEZE2012} . While the dynamics is quite different in mentioned model, e.g. Higgsess are made for  condensation  at classical level, the introduction of "darkon" field is possible as well as in PLQM and the mechanism of the  intergeneration mixing proceeds the same way as in the case of private Higgs models.   Here we avoid further discussion of the mixing and we leave this rather technically involved problem for a future quantitative investigation.

\section{Technicolor input}

Without a detailed specification of an  ETC model we shall mention a crucial points, which are quite general consequences of EWSB  and chiral symmetry breaking in an ETC model.
The current quark mass (i.e the mass when QCD is switched off) appears due to the interaction of SM quarks with Techniquarks at one loop level.
In principle one can estimate $m_q$ from the one loop gap equation (in Euclidean space):   
\be \label{haf}
m_q(p)=\frac{Tr}{4} g^2_E C_E \int \frac{d^4l}{(2\pi)^4} \gamma_{\mu}\frac{M_Q(p-l)}{(p-l)^2+M_Q^2(p-l)}\gamma_{\nu} G^{\mu\nu}_E (l^2) \, ,
\ee
where $g_E$ is ETC coupling, $C_E$ is the ETC group Casimir, $M_Q$ is the techniquark dynamical mass and  $G^{\mu\nu}_E$ is ETC boson propagator with 
ETC mass $M_{ETC}$  larger then Technicolor scale. The Techniquarks receive their masses through the mechanism quite similar to $QCD$, thus
for  Techniquark masses one has
\be
M_Q(p)\simeq \Lambda_{TC} \, \, for \, \, p^2\simeq \Lambda_{TC}^{2}
\ee
which is valid  bellow $\Lambda_{TC}$ due to the walking character of supercritical Technicolor coupling $g_{TC}$.
Above $\Lambda_{TC}$ the dynamical mass is fast decreasing function of the momentum $p^2$:
\be
M_Q(p)\simeq \frac{\Lambda_{TC}}{p^2}\, \, for \, \, p^2\simeq \Lambda_{TC}^{2}
\ee
 Any process, or individual Feynman diagram, which is proportional to fermion masses is softened at large momenta, 
since  for current quark mass one can estimate
\bea \label{jituna}
m_q(p) &\simeq&  \frac{g^2_E}{8\pi^2} C_E  \frac{\Lambda^3}{M_{ETC}^2}\, \, for \, \, p^2<< \Lambda^{2}
\nn \\
m_q(p) &\simeq&  \frac{g^2_E}{8\pi^2} C_E  \frac{\Lambda^3}{M_{ETC}^2} \frac{\Lambda}{p^2}   \, \, for \, \, p^2>> \Lambda^{2} \,
\eea
where the scale $\Lambda$ is identical with $\Lambda_{TC}$ for  $M_{ETC}\simeq \Lambda_{TC}$. In general   $ETC$ and $TC$ scale can be quite distinct, however we simply assume that they can be weakly interacting scalars at $TC$ scale which communicate Techniquarks with the Standard model quarks.  
 Due to this reasons we can simplify and we will use one common scale $\Lambda$  assuming $\Lambda_{TC}\simeq  M$.        
Within this approximation this  TC  mass scale which  characterizes chirality mixing  for Techniquarks as well as the SM quarks.

Three massless technipions are eaten by the longitudinal polarization of $W,Z$, since the Techniquarks couple  minimally to the gauge bosons  $W,Z,\gamma$, 
which gives  rise $W,Z$ bosons masses (without the presence of SM corrections).   While the properties of $Z,W$ bosons are  well known up to the electroweak scale and are subjects of electroweak precision test, a little is known about the boson propagations at TC  scale. Similarly to SM model,  the scattering of low energy longitudinally polarized $W$ and $Z$ bosons is equivalent to the scattering of the Goldstone bosons- technipions in TCD. Like in QCD, the  effective {\it low energy} theory  can be described by Chiral Perturbation Theory, which together with the electroweak oblique corrections, give us rather stringent constrains on the parameter space of TCD \cite{FJS2011}. Not too far above $\Lambda_{TC}$ scale an effective low energy models fail and one should go back into the original microscopical description with the complete decrease of freedom of entire theory.     
Due to the strong coupling a correct treatment of the problem requires  nonperturbative framework, for an examples see for instance \cite{BEN2012,NATALE2013}, noting that  the qualitative knowledge is still quite limited due to  many reasons. The high energy behavior of gauge boson Greens functions, albeit extremely important,  is not numerically known in  theories with dynamical symmetry breaking (like TC or ETC) and  one can make rough estimates only. From the assumption  that $SU(2)\times U(1)$ symmetry should be restored  above TC scale it follows that the gauge boson propagators should be described by their unbroken limits. In covariant Feynman gauge they   read
\be
G_{Z,W,\gamma}^{\mu\nu}(p)\simeq-\frac{g^{\mu\nu}}{p^2}
\ee
valid for $q^2>>\Lambda_{TC}^2$, in contrast to low energy broken symmetry propagator forms which are 
\bea
G^{\mu\nu}_{Z,W}(p)&\simeq&-\frac{g^{\mu\nu}}{p^2-M_{Z,W}^2+\ep}+\frac{\frac{p_{\mu}p_{\nu}}{M_{Z,W}^2}}{p^2-M_{Z,W}^2+\ep} \, ;
\nn \\ 
G^{\mu\nu}_{\gamma}(p)&\simeq&-\frac{g^{\mu\nu}}{p^2} \, 
\eea
presumably valid bellow the Technicolor scale. In words, the polarization functions of all SM gauge bosons must be proportional to  $p^2$ at large Euclidean momenta and the overall prefactor, not shown here, can be absorbed into the usual running of the gauge couplings.

Assuming no parity violations, the full quark propagator reads
\be \label{glumik}
S(p)=\left[A(p)\not p+ B(p)\right]^{-1}
\ee
, where the dynamical mass defined as a ratio of two scalars $M=B/A$   behaves approximately as
\bea  \label{ETCMQO}
m_q(p)&=&Const_1 \,\,\, ; \Lambda_{QCD}^2<<p^2<\Lambda^2 ,
\nn \\
m_q(p)&\simeq&\frac{Const_2*}{p^2} \,\,\, ; p^2>>\Lambda^2 ,
\eea
where the running of $m_q$ at 1 GeV scale is further enhanced by QCD, leading to the additional grow of dynamical quark mass function to the constituent quark values. 

To summarize most important  Technicolor outcome: there exists a scale $\Lambda$ above which $SU(2)_w\times U(1)_Y$ symmetry is restored. The SM quarks  become massless and the longitudinal pieces of gauge boson propagators  go away above this scale. While it can be a multiscale problem in general we will deliberately use single Technicolor scale $\Lambda$ as a common regularization scale for diagrams involving $Z$ bosons and quark mass function in the numerator as well. Note trivially, there are vertices or rather say  "pieces of vertices" which are not regularized by the Technicolor  dynamics, e.g. the renormalization wave function $A$ in Eq. (\ref{glumik}), etc. ..., these, if needed, can be  renormalized. However, no lepton mass renormalization is allowed as the mass terms do not appear at classical Lagrangean. 

\section{Neutrinos mass generation}

Evaluation of the neutrino masses  appears to be easier as the details do not depend so much on the Technicolor dynamics.
Actually , there are no graphs with vector bosons contributing at the lowest order of perturbation theory, which is also the main reason of smallness of neutrino masses.
   
As we will deal with graphs which includes only  weak couplings at the vertices, we will use perturbation theory, regularized at the scale $\Lambda$. The smallness of the coupling causes $A=1+\delta A; \delta A<<1$ , which should be  certainly  valid up to the Technicolor scale. 
In this paper we will neglect changes in  renormalization wave functions $Z=A^{-1}$ and we always take
$A=1$. It allows  to  extract contribution to  the fermion mass from the  selfenergy function  
\bea \label{project}
\delta m &=& \frac{Tr}{4}\Sigma \, .
\eea
Following the discussion in the Section 2, we also require  that the full propagator, which generally reads
\bea
S(p)&=&\not p-\Sigma(p)
\nn \\
&=&\left[ A(p)\not p+ B(p)+ C(p) \not p \gamma _5 + D(p)  \gamma _5 \right]^{-1} \, ,
\eea
has  negligible scalar function $C$ and $D$ (in a sense $C<<1$ and $D<< B$).
The mass function $m$ and physical on-shell mass $m_p$ are related as usually
\bea
m(p)&=& B(p)/A(p)
\nn \\
m(m_p)&=&m_p \, \, ,
\eea 
where in the second line, a real part  of $m$ is understood for unstable muon and tau.

The fountain diagram represent the first nontrivial contribution to the neutrino mass, however it is useful to mention lower order cancellation explicitly.
The one loop  contribution to the neutrino mass reads

\bea
\delta^{1loop} m_{\nu}&=&\frac{y^2}{8\pi^2} m_{\nu} \left(\ln{\frac{\Lambda}{M_s}}-\ln{\frac{\Lambda}{M_p}}\right)
\nn \\
&+&\frac{y^2}{8\pi^2} m_{e} \left(\ln{\frac{\Lambda}{M_s^{\pm}}}-\ln{\frac{\Lambda}{M_p^{\pm}}}\right) \, .
\eea 

Thus for classically massless fermions the one loop  triviality is ensured twice, firstly  it is due to the degeneration of spinless parity partners  $\frac{M_p}{M_s}=1$ (for the four scalars $s={\cal S}_{e,\nu},N^{\pm}$ and for the four pseudoscalars $p={\cal P}_{e,\nu},S^{\pm}$),   secondly the  mass correction is proportional to the mass itself, so it is zero in chiral limit.
 Recall, the  later is valid even nonperturbatively and the masslessness is preserved up to the critical coupling $y^2 \simeq 4\pi^2$ \cite{MIRANSKY}.

Playing few minutes with two loop Feynman diagrams one can show that the cancellation persists  two loops level as well.  
Actually the first nontrivial contribution is given by the sum of fountain type diagram shown in Fig.2. The sum is given by the all  combinations of heavy scalar and  pseudoscalar exchanges at the internal lines. Thank to the traces, there are only sixteen gives nontrivial contributions, those with two $i\gamma_5$ matrices on each individual fermion lines. Each of pseudoscalar (scalar) vertex on the lepton straight line must connect pseudoscalar (scalar) vertex at quark triangle. The others fountains give trivial contribution to the neutrino mass.  Thus the whole three loops contribution can be formally written as
\bea \label{delty}
\delta m_{\nu_e}&=&\delta m_{\nu_e}(e,e;u,d,d)+\delta m_{\nu_e}(e,\nu;u,d,u)
\nn \\
&+&\delta m_{\nu_e}(\nu,\nu;u,u,u)+\delta m_{\nu_e}(\nu,e;u,u,d) \, ;
\eea
where the notation corresponds with the labeling of fermion internal lines in each fountain  diagram. Boson lines were  implicit here, writing them explicitly we can get  for the  first term in the rhs. of Eq. \ref{delty}:
\bea
\delta m_{\nu_e}(e,e;u,d,d)&=&\delta m_{\nu_e}(e,e;u,d,d;s,s,s)+\delta m_{\nu_e}(e,e;u,d,d;p,p,s)
\nn \\
&+&\delta m_{\nu_e}(e,e;u,d,d;s,p,p)+\delta m_{\nu_e}(e,e;u,d,d;p,s,p) \, 
\eea
where $p$ or $s$ means pseudoscalar or scalar with uniquely defined  charges
for a given internal lepton lines content (similar can be written for the remaining three terms). We assume the diagram contribution varies slowly with the square of momentum, thus it is fully legitimate to approximate the physical mass by its infrared value evaluated at $p=0$.

Lets us write explicitly  contribution for the one specific fountain diagram
\bea \label{SSS}
&&\delta m_{\nu_e}(e,e;u,d,d;s,s,s)=
\nn \\
&(-1)&y^6 i\int\frac{d^4k_1}{(2\pi)^4} i\int\frac{d^4k_2}{(2\pi)^4} i\int\frac{d^4l}{(2\pi)^4}
\frac{G(k_1)G(k_2)G(k_1+k_2)}{(l^2-m_u^2+\ep)((l+k_2)^2-m_d^2+\ep)((l-k_1)^2-m_d^2+\ep)}
\nn \\
&&\left\{Tr(\not l+\not k_2+m_d)(\not l+m_u)(\not l-\not k_1+m_d) \, 
\frac{Tr}{4}\frac{\not k_1+m_e}{k_1-m_e^2+\ep}\frac{\not k_2+m_e}{k_2-m_e^2+\ep}
\right.
\nn \\
&+&\left.Tr(-\not l+\not k_1+m_d)(-\not l+m_u)(-\not l-\not k_2+m_d) \, 
\frac{Tr}{4}\frac{\not k_2+m_e}{k_2-m_e^2+\ep}\frac{\not k_1+m_e}{k_1-m_e^2+\ep} \, \, ,
\right\}
\eea
where $G(q)$ is the propagator of spinless heavy boson $G(q)^{-1}=q^2-M^2+\ep$ and where the second Traces over the lepton Dirac matrices are due 
to the projection (\ref{project}). For completeness we mention boson content $(s,s,s)=(N^{\pm},{\cal S},N^{\pm})$ (from the right to the left in the fountain diagram).

We do not write all fountain diagrams explicitly as they uniquely follow from our notation, noting that there are always none or two $\gamma_5$ insertions in the lepton line and corresponding with the same  in the quark line as we are dealing with well parity defined bosons. The example reads
\bea \label{PPS}
&&\delta m_{\nu_e}(e,e;u,d,d;p,p,s)=
\nn \\
&(-1)&y^6 i\int\frac{d^4k_1}{(2\pi)^4} i\int\frac{d^4k_2}{(2\pi)^4} i\int\frac{d^4l}{(2\pi)^4}
\frac{G(k_1)G(k_2)G(k_1+k_2)}{(l^2-m_u^2+\ep)((l+k_2)^2-m_d^2+\ep)((l-k_1)^2-m_d^2+\ep)}
\nn \\
&&\left\{Tr\gamma_5(\not l+\not k_2+m_d)\gamma_5(\not l+m_u)(\not l-\not k_1+m_d) \, 
\frac{Tr}{4}\gamma_5\frac{\not k_1+m_e}{k_1-m_e^2+\ep}\gamma_5\frac{\not k_2+m_e}{k_2-m_e^2+\ep}
\right.
\nn \\
&+&\left.Tr\gamma_5(-\not l+\not k_1+m_d)\gamma_5(-\not l+m_u)(-\not l-\not k_2+m_d) \, 
\frac{Tr}{4}\gamma_5\frac{\not k_2+m_e}\gamma_5{k_2-m_e^2+\ep}\frac{\not k_1+m_e}{k_1-m_e^2+\ep}
\right\}
\eea
with the the overall prefactor dictated by Feynman rules is identical $i^2 i^2=1$.
%


After the trace evaluation one gets for the sum of all fountain diagrams the following expression: 
%
\bea \label{hmota}
\delta m_{\nu_e}&=&32\left\{\int D(udd) \left[ m_u m_d m_d m_e m_e + m_u (l-k_2).(l+k_1) k_1.k_2\right]\right.
\nn \\
&+&\int D(uuu) \left[  m_u^3 m_{\nu_e} m_{\nu_e} + m_u (l-k_2).(l+k_1) k_1.k_2]\right]
\nn \\
&+&\int D(udu) \left[  m_u^3 m_e m_{\nu_e} + m_u (l-k_2).(l+k_1) k_1.k_2]\right]
\nn \\
&+&\left.\int D(uud) \left[  m_u^3 m_e m_{\nu_e} + m_u (l-k_2).(l+k_1) k_1.k_2\right]\right\} \, ,
\eea
where we denote various integral prefactor by the quark content of the triangle subgraph,
explicitly written the definition of the measure reads
\bea
\int D(udd){\cal F}&=&
(-1)y^6 i\int\frac{d^4k_1}{(2\pi)^4} i\int\frac{d^4k_2}{(2\pi)^4} i\int\frac{d^4l}{(2\pi)^4}
\frac{{\cal F}}{(k_1-m_e^2+\ep)(k_2-m_e^2+\ep)}
\nn \\
&&\frac{G(k_1)G(k_2)G(k_1+k_2)}{(l^2-m_u^2+\ep)((l+k_2)^2-m_d^2+\ep)((l-k_1)^2-m_d^2+\ep)}
\eea
From the result it is apparent that the massive quarks circulating in the triangle graph generates effective three boson vertex and in which respect the quark mass hierarchy gives rise the neutrinos mass hierarchy.

We simplify at this stage: We  neglect  all small masses of leptons when they are compared against the large scale   $\Lambda$, the masses of neutrinos are neglected completely (note, there are no potentially large infrared divergence since cutting any two lines of an isolated loop one never meets two massless particles). Furthermore, we use constant quark mass approximation. Nevertheless, one should keep in mind that what actually constitutes the numerical value of virtual triangle subdiagram is the sum of  dynamical quark masses averaged over the product  of the three quark propagator  denominators weighted by the square of momenta characteristic for triangle graphs. This vertex is not a constant, but necessarily run with the external momenta being softer for a large momenta due to the ETC origin of quark masses (\ref{ETCMQO}). The external momenta of triangle  graph become an internal/integration variable of the left and the right  loop of the fountain digram. Clearly one cannot state  only one value at one scale for such vertex, instead of this is the "all scale" momentum behavior which should be seriously taken into account. Happily, this is the "infrared value" which is important for the part of fountains, which remains finite, while this is the large momentum behavior of triangle subgraph,  which drives the value of the part of the fountain which would lead to UV divergence in unregularized fountain otherwise. Later, in accordance with the discussion of preceding section, is the subject of regularization at the Technicolor scale.

In expressions (\ref{PPS}) and (\ref{SSS}) the loop momenta $k_1$ and $k_2$ correspond with the momenta of the left and the right lepton lines respectively. 
Further approximation is based on the observation that this is the large value of the product of lepton momenta $k_1.k_2$ (when compared to $m_l^2$), which makes the  value of the fountains large. 
Thus for a rough estimate of the most important contribution one can consider large $(k_1+k_2)^2$ limit only, which allows exceptionally elegant and simple approximation. The appropriate expression for large $Q^2=(k_1+k_2)^2\simeq 2k_1.k_2$ is derived in the Appendix A.

 As a first let us consider the term proportional to $l^2 k_1.k_2$ which appears in the first line of Eq. (\ref{hmota}).
For this case the effective three scalars vertex is given by the by the expression $\Delta_1(Q;u,d,d)$ of Appendix A. The total result reads
\bea
\frac{1}{(16\pi^2)^3}\frac{\Lambda^2}{2M^2}(-1) L_2 \int dx_0^1 \ln\frac{\Lambda^2(1-x)x+\Lambda^2x}{\Lambda^2(1-x)x+m_d^2x+m_d^2(1-x)}
+\frac{m_u^3}{4M^2 (16\pi^2)^3} L_2 \ln^2\left(\frac{\Lambda^2}{4m_d^2}+\sqrt{1+\frac{\Lambda^2}{4m_d^2}}\right)
\eea
where $L_2=1.172$ corresponds with the numerical value of two loop diagram at zero external momenta,
\be
L_2\simeq \int d^4 l \,\int d^4 k G(l)G(l+k)G(k)\frac{1}{l^2 \, k^2} \, ..
\ee

Relatively straightforward estimate for the first line  term in Eq. (\ref{hmota}) term, which is  proportional to $-(k_1.k_2)^2$ reads
\be
\int D(udd)  m_u (-)(k_1.k_2)^2=
\frac{-1}{8(8\pi^2)^3}\frac{\Lambda^2}{\Lambda^2+M^2}\ln^2{\frac{\Lambda}{M}} 
\ln^2\left(\frac{\Lambda^2}{4m_d^2}+\sqrt{1+\frac{\Lambda^2}{4m_d^2}}\right)
\ee
valid  in large $Q$ limit and  where we also put $(k_1+k_2)^2\simeq \Lambda^2$, which  factorized the rest of the integrand to two separable
log divergent loops with one heavy scalar at internal line. It  leads to factor $\ln^2{\frac{\Lambda}{M}} $ with Pauli-Villars regulator $\Lambda$ . This result is obviously   valid only when   $\Lambda>>M>>m_{q,l}$. For $\Lambda\simeq M$  using  Pauli Villars regularization becomes unjustified and  we rather take 
$(k_1.k_2)^2/(k_1+k^2)^2=\Lambda^2/4$. In this case we can approximately write
\be
\int D(udd)  m_u (-)(k_1.k_2)^2=
\frac{-1}{8(8\pi^2)^3}\frac{\Lambda^2}{M^2}
\ln^2\left(\frac{\Lambda^2}{4m_d^2}+\sqrt{1+\frac{\Lambda^2}{4m_d^2}}\right)
\ee
valid for $\frac{\Lambda^2}{M^2}\simeq 1$.

Using similar formula  the same can be repeated for  the remaining terms in the relation (\ref{hmota}), collecting them together  one arrives to the form for $\delta m$, which has been finally used for the numerical evaluation of electron mass. 
Repeating the same for the muon and tau neutrinos one obtain the analogous formulas, where one need just to replace the appropriate family entries. 

In order to get some first numerical intuition  we take $\Lambda=1 TeV$ as a tentative example. This requires to tune Yukawa coupling in order to have mass sum of the active neutrinos smaller then cosmological constrain 0.5 MeV. Thus within  
 $y=1/400 $, top quark mass $m_t(\Lambda)=150GeV, \, m_b(\Lambda)=4 GeV, m_c(\Lambda)=1GeV, \, m_s(\Lambda)=100 MeV $ ,    we get  
$\delta m_{\nu_e}=1.75 *10^{-5} eV \,  , \,
\delta m_{\nu_{\mu}}= 0.013 eV\,  , \,
\delta m_{\nu_{\tau}}=0.14 eV \, .$
In the next section we will see that for correct  agreement with the charged leptons  one will need to take the scale $\Lambda$  significantly larger  and Yukawa coupling slightly weaker.

\section{Charge lepton mass generation}

 The radiative corrections are responsible for mass generation of the charged leptons as well. However, as the leptons are down components of the electroweak izodoublet, the largest contribution is due to the interaction with the neutral gauge bosons.  Electroweak symmetry breaking gives rise the quark triangle graph, which when jointed to the lepton lines gives rise  the fountain with one heavy particle  and two $Z^0$ or $\gamma$ vector lines presented in. This effect is dramatically enhanced  due to the longitudinal pieces of $Z$ boson propagators.  In fact, this is  almost entirely fountain diagram with two $Z's$ bosons which is responsible for the main contributions to charge lepton masses at Technicolor scale.
To this point, we have evaluated  tau fountain diagram with two photons numerically (using large $k_1.k_2$ approximation), which for reasonable large Technicolor scales  gives the portion which is  $10^{-7}$ times smaller then the contribution from
the fountain with two $Z's$. Actually, when regulated at $\Lambda$ scale, then diagrams including  photon internal lines are all negligible and we neglect them completely from now.

The charged lepton masses are not negligibly small when compared to the quark masses at Technicolor scale. This  makes  a correct estimate more complicated in this case. Roughly pronounced, the selfenergy contribution from quark-lepton fountain diagrams to $\delta_m{_l}$ give  the  current lepton masses at Technicolor scale. Remaining missing contribution to the observed value of lepton masses  is re-induced  from the lower orders  perturbative graphs with nonzero current lepton mass $\delta_m{_l}$ implemented in. Note trivially, mentioned graphs  do not contribute 
in the chiral limit, i.e. when the Technicolor interaction is switched off, however, as we will show explicitly in the next, their contribution as a feedback on non-zero value $\delta_m{_l}$ is relatively large.

In principle, if we would be able to calculate  behavior of quarks and SM gauge bosons Green's functions  all over the Technicolor scale,  then we could be able to obtain the "all scales" lepton dynamical mass function by the solution of  selfconsistent  nonperturbative  quantum equations of motion: the Schwinger-Dyson equations. The system of Schwinger-Dyson equations is a  useful tool in case when perturbation theory fails, e.g. in continuum QCD \cite{A}, and in other strong coupling models \cite{MIRANSKY,ja1,ja2,BEBRSM2009,benes}.   Missing knowledge of the  details of Technicolor dynamics enforce as to use an approximation discussed in the Section II. Other details and approximations made  we describe briefly bellow.

The lepton mass is fully encoded in the lepton selfenergy function  
\be \label{def}
m_l(p)= Tr \Sigma_l(p)
\ee
where the selfenergy $\Sigma$  satisfies  exact Schwinger-Dyson equation, which can be formally written 
\be
\Sigma_l(p)=\sum_{ij} c_{ijl} \int  \frac{d^4 l}{(2\pi)^4} S_i(l)\Gamma_{lji}(l,p) G_{j}(l-p) \Gamma_{ijl} \, , 
\ee  
where all Dirac as well as Lorentz indices are suppressed here for simplicity, $l,i=e,\mu,\tau$ and $j$ stands for all bosons. Note, the higher loops are generated through the expansion of the full trilinear vertices $\Gamma_{lji}(l,p)$ giving us skeleton expansion, while  usual perturbative theory requires further reducible  expansion of the full lepton propagators $S$ and the full boson propagators $G$ as well. 
Vertices $\Gamma_{lji}(l,p)$ satisfy their own Schwinger-Dyson equations which functionally depend on lower and higher vertices as well. We will not solve the set of SDEs numerically, instead we  solve the
Eq. (\ref{def}) semiperturbatively.       
 
For this purpose let us  divide the function $\Sigma$ into two pieces, the first, say $\Sigma_c$ 
let it be the contribution with no  quarks involved and thus giving a trivial contribution in the lepton chiral limit, and the second term -say $\delta m_l$- is generated by the loops with dynamical quarks. 
Because of the weak couplings the first term can be nicely approximated by the sum of one and two loop diagrams
\be
\Sigma_c=\Sigma_{1loop}+\Sigma_{2loops}
\ee
while the lowest order contributing to $\delta m_l$ is just the sum of fountains.

Assuming the fountain diagram has only mild dependence  on $p^2$, it can be estimated by  its infrared value $\delta m(0)$, i.e. the full $\Sigma$ reads
\be   \label{inka}
\Sigma_l(p)=\delta m_l(0)+\Sigma_{l,c}(p) .
\ee
Thus  bellow the Technicolor scale it is equivalent to the $SM$ like expression with nonzero "bare" mass $\delta m$ , where $\Sigma_{l,c}(p)$ is regularized at Technicolor scale.

In this paper we estimate the charged lepton masses from the feedback of the one loop approximation $\Sigma_{1loop}$ with the current lepton mass given by the fountain diagram with two $Z'$. From all three double Z fountains only one contributes in the infrared limit $p\simeq 0$, according to our assumption we neglect two remaining for a while.

Furthermore, we assume a feedback response of $\Sigma_c$  is approximately linear, i.e. the following ratios
\be
\frac{\delta m_l}{\delta m_{l'}}\simeq \frac{m_l}{m_{l'}}
\ee
should hold for all generations.
     
We use usual perturbation theory Feynman graph method to evaluate the all necessary fountain diagrams. 
The appropriate calculation  is straightforward but tedious task and we do not show all intermediate steps in this paper.
To evaluate fountain diagram  we have used some well justified assumption e.g. we have neglected metric tensor part of the gauge boson propagators as
the subdiagrams involving $Z$ propagators are dominated by a a large value of the momenta due to the longitudinal pieces only.
It  allows further simplification, explicitly we have used      
\be
\left<\frac{k_1.k_2}{(k_1+k_2)^2+M^2}\right> \rightarrow \frac{\Lambda^2_{TC}}{4(\Lambda^2_{TC}+M^2)}
\ee
where $k_1$ and $k_2$ are the four-momenta corresponding with  the left and right semi-arc of $Z$ boson lines.

For our special choice $\Lambda_{TC}=M $  the final sum of the dominant contributions reads
\be \label{sauli}
\delta m_l=\frac{32}{2(16\pi^2)^3}\left(\frac{\Lambda^2_{TC}}{M_Z^2}\right)^2 \frac{g^4}{c_w^4}(v_d^2-a_d^2)^2 y^2 m_q 
\ln{\frac{\Lambda^2_{TC}}{m_q^2}}        \, ,
\ee   
 where
\bea
v_d&=&-\frac{1}{4}+\frac{1}{3} s_w^2
\nn \\
a_d&=&-\frac{1}{4}
\nn \\
c_w&=&m_W/m_Z
\eea
and $g=e/s_w$ is $SU(2)_w$ gauge coupling, $e=0.302$, $s_w=sin \theta_W$ for which we take the value at  $M_z=91.2 GeV$ in  MS scheme $s_w^2= 0.231 $.
The  running of the Weinberg angle $\theta_W$ is ignored here. Remind for completeness $v_d,a_d$ are vector and axial vector couplings of $Z$ boson to down type quarks (For upper quarks (and neutrinos) one has $v_u=a_u$ causing that upper quarks loops do not contribute at all).

It is notable that for our choice of the ratio $\Lambda_{TC}/M= 1$ we get the following simple rules
\bea
\frac{\delta_{m_e}}{\delta_{m_\mu}}&=&\frac{m_d}{m_s}\ln\left(\frac{m_s}{m_d}\right)
\nn \\
\frac{\delta_{m_e}}{\delta_{m_\tau}}&=&\frac{m_d}{m_b}\ln\left(\frac{m_b}{m_d}\right) \, ,
\eea
which are independent on Yukawa coupling and depend on the Technicolor scale only through the quark masses $m_q=m_q(\Lambda)$.

The one loop feedback response on emerging nonzero $\delta m$  reads
\be
\Sigma_{c,l}=\frac{g^2}{8\pi^2} \delta_{m_l} (a_d^2-v_d^2) \left(\frac{\Lambda_{TC}}{m_Z}\right)^2 \, .
\ee

If the model is reliable, then fitting our universal Yukawa coupling and Technicolor scale through the relation (\ref{inka}) one should reproduce three masses of the charged lepton known in the nature. To avoid large QCD uncertainties we use the third generation for purpose of our fit. Furthermore, the obtained two parameters, when used to evaluate neutrino masses should leave the neutrino  mass sum under the known cosmological constrain.

To fit the lepton masses one should know the quark masses at $\Lambda_{TC}$, i.e. at the scale where assumed Technicolor interaction becomes important.
 We disregard unknown  renormalization group effects here and we use the masses slightly smaller then their renormalized values at the electroweak scale $M_z$. For this purpose we used $m_{b,\Lambda_{TC}}=2.5 GeV$, recalling here $m_{b}(M_z)=2.87\pm 0.03 GeV$ and we take $m_{t,\Lambda}=150 GeV$ recalling  here $m_{t}(M_z)=172\pm 3 GeV$. 
Within the  Yukawa coupling value $y=0.042 (y^2=1/550)$ and the scale $\Lambda_{TC}=15 TeV$, herein identical with the masses of all fundamental private 
$SU(2)$ spinless bosons, we get for the third generation of the leptons the following values:
%
\be
  m_{\tau}=(0.19+1.06) GeV=1.25 GeV \, \, ;\, \, 
 \delta m_{\nu_{\tau}}=0.42 eV \, ,
\ee
where the first number corresponds with the contribution from the fountain diagram alone, i.e.  $\delta_m$ and  the second value corresponds with the one loop response with single $Z$ boson line. 

Taking the mass values of the first generation quarks as following $m_{u,\Lambda_{TC}}=0.4; m_{d,\Lambda_{TC}}=0.5 MeV $, 
we get almost our world electron and reasonably light electron neutrino
\be
m_{e}=(73+421)keV=494 keV  \, \, ;\, \,
\delta m_{\nu_{e}}=1.1* 10^{-5} eV
\ee
Within the second generation parameters $m_{c,\Lambda_{TC}}= 0.5 GeV, m_{s,\Lambda_{TC}}= 50 MeV $
we get for the second generation
\be
  m_{\mu}=(5+31)MeV=36 MeV  \, \, ;\, \,
 \delta m_{\nu_{\mu}}=5.6 *10^{-3} eV.
\ee

\section{Down Quarks}

Although we assume the heavy quarks acquire their masses by some sort of ETC dynamics, the mechanism 
described above bring a non-negligible contribution to down quarks as well. 
Up to the relatively small contribution from  fountain diagrams with two gluons \cite{future}, then the contributions to down type quark masses are the same as in the case charged leptons. This is the consequence of the universality of the Yukawa  and neutral electroweak current.
Switching off gluons we simply get
\be \label{fun}
 \hat{m}_b(\Lambda)=m_{\tau} \,\, ;\,\, \hat{m}_c(\Lambda)=m_{\mu}\,\, ; \,\, \hat{m}_d(\Lambda)=m_{e}.
\ee
where hats indicate the "bare" SM quark mass, i.e. the the mass without QCD loops. The evolution down to the QCD scale would require 
more serious solution of QCD gap equation. However, it is already now clear that already now there is a littler space for additional 
contribution stemming from  an ETC dynamics. From this is clear, that if model taken seriously, these are almost only upper quarks which 
feel Techniquarks through ETC, which should violate weak izospin symmetry to get observed up-down mass asymmetry.
In this sense the Eq. (\ref{fun}) represents direct quantitative  constrain on the ETC part of the entire, not yet completely pronounced model.

\section{Conclusion and discussion}

We have proposed the hybrid model  for EWSB and fermion mass generation. The model  is based on partial utilization of heavy scalar doublets and partially on the  ETC mechanism, the later turns to be working in the quark sector only, e.g. it is highly izospin asymmetric with possibility that  ETC can be solely reduced to the Top assisted Technicolor.
The lepton masses are generated purely radiatively. The mass is catalyzed  through the loops diagram with closed quark loops, which firstly appear at  3 loops level. Once the small mass term is seeded, its contribution if further enhanced by the lower  diagrams. This feedback is missing in the case of neutrinos, which is  a consequence of the form of the Standard Model neutral current.    

Comparing the standard ETC scenario, the extension of the SM is actually quite minimal, e.g. the all Yukawas and the all scalar masses are universal, further unified with Technicolor scale.  This universal Yukawa coupling is relatively small $\alpha_y=10^-3-10^{-4}$, while the Technicolor scale has been found relatively large $\Lambda \simeq 12 TeV$. Surprisingly these two numerical parameters are enough to feed up all lepton and down quark  masses.
The model is surprising simple and makes the model tentative for a future less approximative studies.

The model is not UV complete, however the problem of naturalness has been moved to much higher scale $\simeq 100 TeV$ (assuming there is no fine tuning).
It is notable, that in this case, UV completion can be accomplished by quantum gravity in its non-supersymmetric extension \cite{WUZH2008}.      
The mixing has not yet been incorporated,  however already now it is natural to expect that the normal neutrino hierarchy is realized. 
Clearly two much lighter states $\nu_1$ and $\nu_2$ could be  preferably mixed from electron and muon neutrino, the heaviest neutrino mass eigenstate could receive dominant part from $\nu_{\tau}$. Imposing a discrete symmetry on an (not to much) extension of presented  model here, one could incorporate neutrino mixing   as well  (for review of such possibilities see \cite{KILU2013}). This can accomplished in a way that no pathology arise in the model.

On the other side , one can expect some quantitative changes due to the effect not considered here. 
In the beginning we have shown the electroweak oblique corrections $S,T$ remains intact  due to the assumed degeneracy of izodoublet components.
We expect this is  valid when loop corrections are taken into account and the contributions to $S,T$ parameters  stay invisibly small.   
On the other side, even small degeneracy between the scalar and the pseudoscalar mass eigenvalue components leads to a
new contribution to neutrino masses at two loop level. In fact this is proportional to $\Delta S\simeq ln{\frac{M_u}{M_d}}$. We expect dismantling of degeneracy is small and can be compensated by a small readjustment of the Yukawa coupling and heavy scalar mass.      

We did not consider a possible momentum flow in the fountain diagrams and we omitted diagrams in Fig.{ref{fig3}} since they do not contribute at zero momenta. While a more accurate calculation remains to be done, let us assume for a while that each  omitted diagram contributes by the same amount as the  diagram already considered. The stability ensures the physics does not change too much, which reduces the  Technicolor scale to the value $\Lambda=12 TeV$, while  keeping the masses of the  all leptons almost unchanged (keeping the same Yukawa).

We have neglected an assumed small contributions. The examples are  purely QED selfenergy corrections.
They should be small if the  regularization  mimic reasonably  unknown Green's function behavior. This is generally true in vectorlike {\it asymptotic free} gauge theory which posses dynamical chiral symmetry breaking. However, due to a  bit complicated chiral symmetry breaking mechanism  here, and one has to actually prove that dynamical lepton mass function decreases fast enough above Technicolor scale. This remains to be done.      

Observed 126 GeV Higgs is assumed to be a Technisigma in our model, i.e. the lightest bound state made from the  Techniquarks.
Such Technisigma interacts with leptons quite indirectly and it is mediated through the radiative corrections.
These corrections has similar origin  as the lepton masses itself, thus very naturally one can expect the decay rates to leptons can be SM-like, however it can be distinguished quantitatively.  Thus the LHC 126 scalar decay rate to $\tau^{+}\tau^{-}$ can potentially distinguish between SM Higgs and other scenaria like the one  suggested in this paper. After a possible improvement \cite{GNNSW} of the method and discrimination of the signal background stemming such as $Z_0\rightarrow\tau,\tau$ the  LHC 126 scalar decay rate to $\tau^{+}\tau^{-}$ can be measured with reasonable accuracy.

\appendix
\section{Approximative formulas for triangle graphs}

This is the large overlap of the left and the right loop momenta in the fountain diagram which gives desired vast amount of neutrino mass
,here  we derive  a simple expression valid for quark  triangle graphs in the large momentum limit. Let us denote various external  momenta of the triangle, $k_1$ and $k_2$ and $k_1+k2$ in a way that when the triangle is matched into the fountain the momentum $k_1+k_2$ belongs to the middle narrow heavy boson line.  In the all cases, the important of quantity  which gives is important for the size chiral breaking generated radiatively  is the product of virtual momenta $k_1.k_2$, corresponding  thus to the large momentum limit $Q^2=(k_1+k_2)^2\simeq \Lambda^2 >>m_q$ of the quark triangle subdiagram.
For simplicity we take all quark masses equal here, noting the approximation we make for different flavour at the end. The evaluation is straightforward and proceeds usual way, i.e.textbook Feynman parameterization, Wick rotation and Euclidean momentum integration. For completeness  we write down some of intermediate steps for each momentum integral separately.

The dominant integral  in the quark triangle graph is the following  
\be
\Delta_1(Q;i,j,l)= i\int\frac{d^4k_1}{(2\pi)^4}
\frac{(-1)m_{q_i} l^2}{(l^2-m_{q_i}+\ep)((l+k_2)^2-m_{q_j}^2+\ep)((l-k_1)^2-m_{q_l}^2+\ep)} \, ,
\ee
where we included the numerator quark mass in order to indicate where the regularization originates. Using $\Lambda $ as a suitable cut-off
we get    
\be
\Delta_1(Q;i,j,l)= -m_{q_i}^2 m_{q_i}\Delta_0(Q;i,j,l)+\frac{(-m_{q_i})}{16\pi^2}\int_0^1 dx \ln \frac{(k_1+k_2)^2x(1-x)+\Lambda^2}
{(k_1+k_2)^2x(1-x)+m^2_{q_j}x+m^2_{q_l}(1-x)} \, ,
\ee
where $i,j,l$ are flavour indices.  Important prefactor quark mass $m$ is  written here, telling us explicitly hat the regularization  at the Technicolor scale $\Lambda$ is introduced.
  
The function $\Delta_0$ is finite and at this stage it  does not need any regularization as it  corresponds with the  
the following scalar momentum integral   
\bea \label{comilator}
\Delta_0(Q)&=& i\int\frac{d^4k_1}{(2\pi)^4}
\frac{(-1)}{(l^2-m_q+\ep)((l+k_2)^2-m_q^2+\ep)((l-k_1)^2-m_q^2+\ep)}
\nn \\
&=&\frac{1}{(4\pi)^2}\int_0^1 dy\int_0^y dx \frac{1}{-k_2^2 y(1-y)-k_1^2 x(1-x)-2k_1.k_2 x(1-y)+m_q^2}
\nn \\
&\simeq&\frac{1}{(4\pi)^2}\int_0^1 dy\int_0^y dx \frac{-1}{(k_1+k_2)^2 x(1-y)-m_q^2}
\nn \\
&=&\frac{-1}{(4\pi)^2}\int_0^1 dz \frac{\ln{\left(1-\frac{k_1+k_2)^2 z(1-z)}{m_q^2}\right)}}{(k_1+k_2)^2 z}
\eea 
where we have considered all the quark lines belong to the same flavour.
For spacelike momentum $Q=k_{E1}+k_{E2}$, $Q^2>0$ one can integrate over the variable $z$, giving as simple expression
\be
\Delta_0(Q)=\frac{2}{Q^2(4\pi)^2}{\mbox arcsinh}^2(Q/2m)
\ee
or equivalently   
\be
\Delta_0(Q)=\frac{1}{2\, Q^2 \,(4\pi)^2}
\ln^2{\left(\frac{Q^2}{4m_q^2}+(1+\frac{Q^2}{4m_q^2})^{1/2}\right)} 
\ee

If the triangle is connected with the lepton line via an exchange of electrically charged boson, then the masses of entering  quarks are different. In this case the resulting formula  slightly complicates. However, numerically it appears quite adequate to replace the threshold of identical quarks $4m_q^2$ in Eq. (\ref{comilator}) by the sum of the quark masses entering the vertex , i.e. by $t^2=(m_{qi}+m_{qj})^2$
where quark lines $i$ and $j$  connect the vertex with the external momentum $k_1+k_2$ .  Thus the approximative formula reads
\be
\Delta_0(Q)=\frac{2}{Q^2(4\pi)^2}{\mbox arcsinh}^2(Q/t)
\ee
where $t=m_u+m_d$, when calculating electron neutrino mass, and $t\simeq m_c$ and $t\simeq m_t$ for another two possible combinations of inequivalent flavour combinations contributing in the cases of muon and tau neutrinos.

\end{document}